\documentclass[12pt]{iopart}

\usepackage{hyperref}
\usepackage{amssymb}
\usepackage[final]{graphicx}

\begin{document}

\rightline{SPIN-06-37, ITP-UU-06-47} \rightline{}

\title[Problems and hopes in nonsymmetric gravity]{Problems and hopes in nonsymmetric gravity\footnote{As presented by the authors at the 2nd International
Conference on Quantum Theories and Renormalization Group in
Gravity and Cosmology (IRGAC) 2006, Barcelona.}}

\author{Tomas Janssen$^1$
        and Tomislav Prokopec$^2$}
\address{Institute for Theoretical Physics, University of
Utrecht, Leuvenlaan 4, Postbus 80.195, 3508 TD Utrecht, The
Netherlands}
\eads{\mailto{$^1$T.Janssen2@phys.uu.nl},\mailto{$^2$T.Prokopec@phys.uu.nl}}

\begin{abstract}
We consider the linearized nonsymmetric theory of gravitation
(NGT) within the background of an expanding universe and near a
Schwarzschild mass. We show that the theory always develops
instabilities unless the linearized nonsymmetric lagrangian
reduces to a particular simple form. This form contains a gauge
invariant kinetic term, a mass term for the antisymmetric
metric-field and a coupling with the Ricci curvature scalar. This
form cannot be obtained within NGT. Based on the linearized
lagrangian we know to be stable, we consider the generation and
evolution of quantum fluctuations of the antisymmetric
gravitational field ($B$-field) from inflation up to the present
day. We find that a $B$-field with a mass $m\propto
0.03(H_I/10^{13}~{\rm GeV} )^4 \rm{eV}$ is an excellent dark
matter candidate.
\end{abstract}
 \submitto{J. Phys. A: Special
Issue IRGAC 2006}

\maketitle
\section{Introduction}

While Einstein's general relativity  (GR) has stood all direct
experimental tests~\cite{Will:2005va}, there are also reasons to
try to extend GR. For example
 the mysterious nature of dark energy and dark matter might become
resolved within a modified theory of gravity.

 Another reason to try to extend GR is the notion of generality.
Within the framework of GR torsion is not included in a natural,
geometric way. Indeed any calculation of the connection (either by
requiring metric compatibility, or by using the first order
formalism) leads to the (symmetric) Levi-Civit\`a connection. One
is then free to add torsion, but torsion does not follows
naturally from the theory. An interesting generalization of GR
would generate torsion in a purely geometric way, analogous to the
way the Levi-Civit\`a connection is generated in GR.

 The Nonsymmetric Gravitational Theory (NGT)~\cite{Moffat:1994hv}
is an extension of GR that drops the standard axiom of GR that the
metric is a symmetric tensor. Thus we decompose the general,
nonsymmetric metric $g_{\mu\nu}$ in it's symmetric and
antisymmetric parts
\begin{equation} \label{naive}
g_{\mu\nu}=G_{\mu\nu}+B_{\mu\nu},
\end{equation}
where $G_{\mu\nu}=g_{(\mu\nu)}$, $B_{\mu\nu}=g_{[\mu\nu]}$ and
$(\cdot)$ and $[\cdot]$ indicate normalized symmetrization and
anti-symmetrization, respectively. Indeed there is no physical
principle that tells us that the metric should be symmetric and
therefore such a generalization is very interesting to study.

 Indeed the extra structure of NGT produces
interesting results on the issues of dark energy and dark
matter~\cite{Moffat:2004bm}~\cite{Moffat:2004nw}~\cite{Prokopec:2006kr}
and it will also be clear that such a theory produces torsion in a
very natural way. Unfortunately the nonsymmetric theory of
gravitation suffers from all kinds of problems. The first main
problems is the non-uniqueness of the theory, as described
in~\cite{Janssen:2006nn}. Since torsion is available and since the
linearization procedure is not unambiguous, the final linearized
lagrangian is (degenerately) determined by 11 free parameters. The
second problem, as described in~\cite{Damour:1992bt}, is the
possibility of propagating ghost modes.  Fortunately this problem
can be relatively easy solved by the introduction of a mass term
for the $B$-field~\cite{Moffat:1994hv}~\cite{Clayton:1995yi}.

    In this talk we consider NGT linearized around a GR
configuration. By explicitly constructing two different
backgrounds (FLRW-universe and Schwarzschild) we show that the
evolution of the $B$-field is unstable. By considering the most
general form of the linearized lagrangian, we can explicitly point
out which terms cause these instabilities.
In~\cite{Janssen:2006nn} it is both shown that these terms cannot
be removed and that these terms are not a relic of the
linearization.
    Based on this analysis we are able to
write down a consistent, stable linearized lagrangian for the
B-field. We next canonically quantize the B-field in inflation and
follow its dynamics in radiation and matter era. This analysis
shows that the B-field is an excellent dark matter candidate,
provided the mass is of the order of the neutrino masses.

\section{The linearized Lagrangian} \label{slin}
Since GR is very successful, it is natural to assume that any
modification of the theory should be relatively small. Therefore
we consider NGT in the limit of a small $B$, but an arbitrary $G$.
The linearization of the full, general lagrangian is done in
Appendix A of~\cite{Janssen:2006nn}. The result is
\begin{eqnarray} \label{lagrangian}
    \mathcal{L}=\sqrt{-G}\bigg[&R+2\Lambda-\frac{1}{12}H^2+(\frac{1}{4}m^2+\beta R)B^2\\
    \nonumber&-\alpha R_{\mu\nu}B^{\mu\alpha}B_\alpha{}^\nu-\gamma R_{\mu\alpha\nu\beta}B^{\mu\nu}B^{\alpha\beta}\bigg]+\mathcal{O}(B^3).
\end{eqnarray}

Here the curvature terms $R_{\mu\alpha\nu\beta}$, $R_{\mu\nu}$ and
$R$ all refer to the background, GR, curvature. $H_{\mu\nu\rho}$
is the field strength associated with $B_{\mu\nu}$. The
coefficients $\alpha,\beta$ and $\gamma$ are determined by the
parameters of the 'full' lagrangian and the unambiguous
decomposition of the metric in its symmetric and anti-symmetric
parts. It is important to note that one cannot consistently choose
the parameters of the full theory in such a way that $\gamma=0$
(see appendix A of~\cite{Janssen:2006nn}). The parameters $\alpha$
and $\beta$ can in principle be set to zero, however a priori
there is no reason to do this. A mass is naturally generated in
the presence of a nonzero cosmological constant and in fact one
has
\begin{equation} \label{assumption}
\frac{1}{4}m^2=\Lambda\Big(\frac{1}{2}-\rho+4\sigma\Big)\propto
10^{-84}~{\rm GeV}^2.
\end{equation}
where we assume that the parameters $\rho$ and $\sigma$ are order
unity. Note that the inequality is not necessarily true at all
times, since the cosmological term may change during the evolution
of the Universe (for example during phase transitions). The field
equations derived from the lagrangian (\ref{lagrangian}) are
\begin{eqnarray} \label{fieldeq}
&(\sqrt{-G})^{-1}\frac{1}{2}\partial_\rho
(\sqrt{-G}H^{\rho\mu\nu})
 + (\frac{1}{2}m^2+2\beta R)B^{\mu\nu}\\
\nonumber&\qquad\qquad -\alpha(B^{\nu\alpha}R^\mu{}_\alpha
  +B^{\alpha\mu}R^\nu{}_\alpha)-2\gamma
B^{\alpha\beta}R^\mu{}_\alpha{}^\nu{}_\beta +\mathcal{O}(B^2)=0\\
&R_{\mu\nu}-\frac12 R G_{\mu\nu}-\Lambda G_{\mu\nu}+\mathcal{O}(B^2)=0
\,.
\end{eqnarray}
We see that to this order the field equations decouple and it
makes sense to consider the symmetric background, to be just a GR
background. The theory then reduces to an antisymmetric tensor
field coupled to GR.

\section{Instabilities in NGT}
 We first focus on the dynamics of the B-field in an expanding universe\footnote{This section is based
on Ref.~\cite{Janssen:2006nn}}~\cite{Prokopec:2005fb}. Our
background metric is given by the (conformal)
Friedmann-Lemaitre-Robertson-Walker metric (FLRW):
\begin{equation} \label{metric}
    G_{\mu\nu}=a(\eta)^2\eta_{\mu\nu},
\end{equation}
where $\eta_{\mu\nu}=\rm{diag}(1,-1,-1,-1)$, $\eta$ is conformal
time and $a(\eta)$ is the conformal scale factor. The conformal
time is related to the standard cosmological time by, $ad\eta=dt$.
The scale factor during the different cosmological eras is given
in table \ref{tabelletje}, where $H_I \sim 10^{13}~{\rm GeV}$ is
the Hubble parameter during inflation and $\eta_{eq}$ is the
conformal time at matter-radiation equality.
\begin{table}
\caption{The scale factor and conformal time in different eras}
\label{tabelletje}
\begin{center}
\begin{tabular}{|c|c|c|}
  \hline
  era & $a$ & $\eta$ \\
  \hline
  de Sitter inflation & $a=-\frac{1}{H_I\eta}$ &$\eta\leq-\frac{1}{H_I}$\\
  Radiation & $a=H_I\eta$ & $\frac{1}{H_I}\leq\eta\leq\eta_{eq}$\\
  Matter & $a=\frac{H_I}{4\eta_{eq}^2}(\eta+\eta_{eq})^2$ &$\eta \geq \eta_{eq}$\\
  \hline
\end{tabular}
\end{center}
\end{table}

For the following discussion we focus on the 'electric' mode of
the B-field: $E_i\equiv B_{0i}$. (the 'magnetic' mode turns out
not to be very interesting for our present purpose). If we
evaluate the lagrangian (\ref{lagrangian}) and the field equations
(\ref{fieldeq}) in the FLRW background we find the follwing
equation of motion
\begin{equation} \label{finaleom}
\bigg[\partial_0\partial_0-\frac{\mathcal{Y}}{\mathcal{X}}\delta^{ij}\partial_i\partial_j+M^2_{eff}\bigg]\tilde{E}=0,
\end{equation}
where
\begin{equation} \label{transformation}
    {E}=\frac{\sqrt{\mathcal{Y}}}{\mathcal{X}}\tilde E,
\end{equation}
and the effective mass term is given by
\begin{equation}\label{effmass}
M^2_{eff}=-2\mathcal{Y}a^2+\frac{\mathcal{Y}''}{2\mathcal{Y}}-\frac{3(\mathcal{Y}')^2}{4\mathcal{Y}^2}
\,.
\end{equation}
Furthermore we have defined
\begin{eqnarray}
&\mathcal{X}=a^{-2}\bigg((12\beta+2\alpha)\mathcal{H}^2+(12\beta+4\alpha-2\gamma)\mathcal{H}'-\frac{1}{2}m^2a^2\bigg) \label{X}\\
&\mathcal{Y}=a^{-2}\bigg((12\beta+4\alpha-2\gamma)\mathcal{H}^2+(12\beta+2\alpha)\mathcal{H}'-\frac{1}{2}m^2a^2\bigg)
\label{Y}
\end{eqnarray}
and
\begin{equation}
    \mathcal{H}=\frac{a'}{a},
\end{equation}
where a prime indicates a derivative with respect to conformal
time. We see from (\ref{finaleom}) that $\tilde{E}$ behaves just
as a massive vector field, \emph{as long as,
${\mathcal{Y}}/{\mathcal{X}}>0$}. On the other hand, if
${\mathcal{Y}}/{\mathcal{X}}<0$ we see that the spatial
derivatives appear with the 'wrong' sign. Since in fourier space
these derivatives generate a term proportional to minus the
momentum squared, we see that a wrong sign will lead to an
exponential growth of the field. Large momenta are no longer
suppressed and thus the field will grow without bounds. One could
worry about the cases when, $M^2_{eff}<0$. However on dimensional
grounds, the effective mass squared scales in the worst case as,
${1}/{\eta^2}$. Such a scaling results in a standard power-law
enhancement on super-Hubble scales~\cite{Prokopec:2005fb} and
presents no problem.

\subsection{Instabilities during Radiation era}
In de Sitter inflation ${\mathcal{Y}}/{\mathcal{X}}=1$, and thus
the field dynamics are completely regular. However during
radiation era we obtain
\begin{equation}
 \bigg[\partial_0\partial_0
    -  \frac{H_I^2m^2\eta^4+4(\gamma-\alpha)}
           {H_I^2m^2\eta^4-4(\gamma-\alpha)}\delta^{ij}\partial_i\partial_j
    +  M^2_{r}
 \bigg]\tilde{E}_r=0.
\label{EOM:radiation era}
\end{equation}
Here $M_{r}$ is the effective mass during radiation, whose precise
form is not important for us. We see however, that we might have
problems with the sign of the coefficient in front of the spatial
derivatives. For example if we look at the beginning of radiation
era ($\eta={1}/{H_I}$) we see that if we want
${\mathcal{Y}}/{\mathcal{X}}$ to be positive, we need that
${m^2}/{H_I^2}$ is \emph{at least}, $\mathcal{O}(\alpha-\gamma)$.
In other words we approximately need:
\begin{equation} \label{error}
m\geq |\alpha-\gamma|H_I\sim |\alpha-\gamma|\times 10^{13}~{\rm GeV}
\,,
\end{equation}
which, unless $|\alpha-\gamma|$ is very small, contradicts
Eq.~(\ref{assumption}). Therefore if we require
${\mathcal{Y}}/{\mathcal{X}}$ to be positive, we could drop the
purely geometric origin of the lagrangian and add by hand a large
($10^{13}~{\rm GeV}$) mass for the $B$-field, we could fine-tune
$\alpha$ or $\gamma$ such that $\alpha-\gamma$ is sufficiently
small to satisfy the bound~(\ref{error}), or we could use the more
natural requirement that $\alpha=\gamma$. On theoretical grounds
only the last of these solutions is satisfactory. A big problem
with the first solution is that, while we can always find a mass
where the evolution of the mode is stable, we can than also think
of more extreme situations where the mode once again becomes
unstable. Therefore we conclude that a natural theory should have
$\alpha=\gamma$.

We have also investigated matter era and power-law inflation and
we find that similar instabilities are present. However also in
these cases $\alpha=\gamma$ stabilizes the system.
\subsection{Instabilities around a Schwarzschild mass}
We have done a similar analysis in a Schwarzschild background. We
won't give any details here (see section 4
of~\cite{Janssen:2006nn}), but will only mention that similar
instabilities are present; however now the requirements for a
stable system are either
\begin{equation}
    \gamma=0
\end{equation}
or
\begin{equation} \label{massobey}
    m^2>\frac{4\gamma G_N\hbar^2}{c^4}
    \frac{M_0}{r_0^3}\qquad\qquad [kg^2],
\end{equation}
where we explicitly plugged back factors of $c$, $h$ and $G_N$.
$M_0$ is the mass of the object we are considering and $r_0$ is
the distance where we require stability. for $\gamma$ order 1 this
requires e.g. for the exterior of a neutron star ($M_0 \propto
M_{\rm{sun}}$ and $r_0\propto 20\rm{km}$):
\begin{equation}
    m\gtrsim\sqrt{|\gamma|}\times 10^{-19}~{\rm GeV}
\end{equation}

However, on theoretical grounds, it is more appealing to require
that the $B$-field is stable for all values of $M_0$ and $r_0$.
This can only be achieved if we choose $\gamma=0$. However as
noted in section 2, this choice is not possible within our
linearization of NGT.

\section{Antisymmetric metric field as Dark matter}
\begin{figure}
\begin{center}
\includegraphics[width=3.5in]{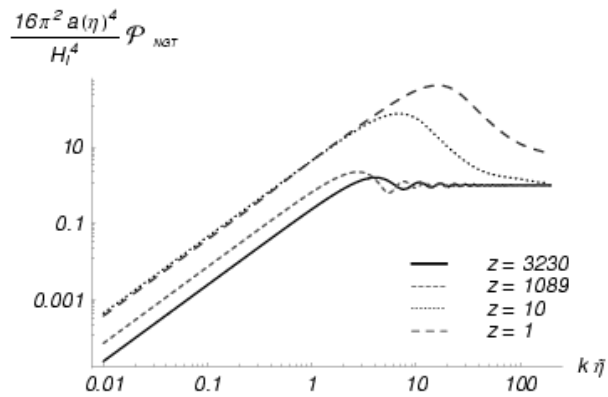}
\caption{Snapshot of the power spectrum for $m H_I
\eta_{eq}^2=10^{-2}$} \label{powerfigure}
\end{center}
\end{figure}

Based on the previous section we know that the only consistent
linearized lagrangian for the $B$-field is
\begin{equation} \label{lagrangiangood}
    \mathcal{L}=\sqrt{-G}\bigg[R+2\Lambda-\frac{1}{12}H^2+(\frac{1}{4}m^2+\beta
    R)B^2\bigg].
\end{equation}
While this lagrangian is not obtainable in NGT, we like to stress
that our linearization procedure of NGT lacks any guiding
principle (which is reflected in the non-uniqueness of the
theory). The analysis above shows that if we want to make sense of
nonsymmetric gravity we need to find a guiding principle that,
upon linearization, leads to (\ref{lagrangiangood}). For now we do
not know this principle, but we can still study
(\ref{lagrangiangood}).
    In this section\footnote{Based on
Ref.~\cite{Valkenburg}} we consider the generation and evolution
throughout the cosmological history of quantum fluctuations of the
$B$-field. In particular we only consider the longitudinal degrees
of freedom of the 'magnetic' component
~\cite{Prokopec:2006kr}~\cite{Prokopec:2005fb}~\cite{Valkenburg}
\begin{equation}
    B_{ij}\equiv-\epsilon_{ijk}B_k,
\end{equation}
since this mode gives the dominant contribution to the energy
density in the limit $m\rightarrow 0$. For simplicity we take
$\beta=0$, but keep the mass arbitrary. When compared to
(\ref{assumption}) this means we allow the presence of a small
bare mass for the $B$-field. In order to quantize the field we
perform a Fourier transformation
\begin{equation}
B^L(x)=\int\frac{d^3k}{(2\pi)^{3/2}}\Bigg[e^{i\vec{k}\cdot\vec{x}}B^L(\eta,\vec{k})b_{\vec{k}}+e^{-i\vec{k}\cdot\vec{x}}B^{L\star}(\eta,\vec{k})b^\dagger_{\vec{k}}\Bigg],
\end{equation}
where $\eta$ is once again conformal time as given in table
\ref{tabelletje}, with canonical commutation relations
\begin{equation}
[b_{\vec{k}},b^\dagger_{\vec{k}'}]=(2\pi)^3\delta^3(\vec{k}-\vec{k}')
\end{equation} During de Sitter inflation we find that the mode
functions approach the conformal vacuum
\begin{equation}
B^L_{\rm{inf}}\propto
\frac{1}{\sqrt{2k}}e^{-ik\eta}+\mathcal{O}\bigg(\frac{m^2}{H_I^2}\bigg).
\end{equation}
During radiation era the field equations are solved by
\begin{equation}
B^L_{\rm{rad}}=\frac{1}{\sqrt{2k}}\Bigg[\alpha_{\vec{k}}\Bigg(1-\frac{i}{k\eta}\Bigg)e^{-ik\eta}+\beta_{\vec{k}}\Bigg(1+\frac{i}{k\eta}\Bigg)e^{ik\eta}\Bigg]+\mathcal{O}\bigg(\frac{m^2}{H_I^2}\bigg)
\end{equation}
with the Wronskian condition that
\begin{equation}
    |\alpha_{\vec{k}}|^2-|\beta_{\vec{k}}|^2=1
\end{equation}
and we choose $\alpha$ and $\beta$ such that the solutions match
at the inflation-radiation transition. Unfortunately we cannot
analytically solve the equations of motion in matter era, so there
we need to use numerical analysis.
    We are interested in the power spectrum, which is given by~\cite{Prokopec:2006kr}
\begin{equation}
    P_B(\vec{k},\eta)=\frac{H_I^4}{4\pi^2 a^4}\Bigg[|\partial_\eta
    B^L_{\vec{k}}(\eta)+\frac{a'}{a}B^L_{\vec{k}}(\eta)|^2+(k^2+a^2m^2)|B^L_{\vec{k}}|^2\Bigg].
\end{equation}
A snapshot of this power spectrum, during matter era, for
different redshifts is given in figure \ref{powerfigure}.
We find that at late times the power spectrum becomes dominated by
a characteristic peak. This peak is caused by modes that are
superhorizon ($k\eta\lesssim 1$) at equality ($z=3230$), but start
to scale as nonrelativistic matter ($\propto a^{-3}$)in matter era
and enter the horizon. Modes on small enough scales
($k\eta>a/a_{eq}$) are effectively massless and scale as
relativistic matter $\propto a^{-4}$. The position of the peak is
determined by the mass of the $B$-field. In fact we have
\begin{equation}
    k_{\rm{peak}}=\sqrt{H_I m}
\end{equation}
    Now that we know the power spectrum, we can calculate the
energy density of the $B$-field, defined by
\begin{equation}
    \rho_B=\int\frac{dk}{k}
    P_B.
\end{equation}
A good dark matter candidate should have an energy density
\begin{equation}
    \frac{\rho_B}{\rho_{\rm{rad}}}=1\qquad\qquad\rm{at}\qquad\qquad
    \eta=\eta_{eq},
\end{equation}
where $\rho_{\rm{rad}}$ is the enrgy density of the cosmic
radiation. The calculation is done in~\cite{Valkenburg} and it is
found that
\begin{equation}
    m   =2.8\times
    10^{-2}\Bigg(\frac{10^{13}\rm{GeV}}{H_I}\Bigg)^4\rm{eV}
\end{equation}
gives the right energy density.

\section{Discussion and conclusion}
We have shown that, while the nonsymmetric theory of gravitation
is an extremely interesting extension of general relativity to
study, the modes of the antisymmetric metric field are unstable.
This instability manifests itself through a wrong sign in front of
spatial derivatives in the equations of motion. Such a wrong sign
means that large momenta are no longer suppressed, and therefore
the field grows without bounds. We showed that the troublesome
terms in the lagrangian (\ref{lagrangian}) are the coupling to the
Riemann tensor and the Ricci tensor. Furthermore
in~\cite{Janssen:2006nn} it was shown
 that the first of these terms cannot be
removed in NGT and that the instabilities are not a relic of the
linearization.
    However, our linearization procedure was rather naive, and it
lacks a good guiding principle. Our analysis shows that \emph{if}
one could find a good principle from which to construct a
nonsymmetric theory of gravitation (e.g. by considering complex
manifolds as in ~\cite{Chamseddine:2005at} ~\cite{Mann:1982}), the
linearized lagrangian \emph{must} have the form of
(\ref{lagrangiangood}).
    Based on this knowledge, we've studied the evolution of quantum fluctuations, generated at
    inflation,
throughout the cosmological history. We find that the $B$-field
has the right energy density to fully take account for the dark
matter energy density if the mass of the field is given by
$m=2.8\times
10^{-2}\Bigg(\frac{10^{13}\rm{GeV}}{H_I}\Bigg)^4\rm{eV}$.
Furthermore the power spectrum develops a characteristic peak,
that for this mass and $z=10$ (start of structure formation) has a
length scale coincidentally corresponding to the earth sun
distance.
    Although the mass of the $B$-field is small (equivalent to the mass of the $\tau$-neutrino), it still
is \emph{cold} dark matter. Indeed, since the field does not
couple to matter fields, it cannot thermalize and therefore the
spectrum stays primordial and highly non-thermal. Because of this,
it does not suffer from the problems that neutrino dark matter
has.
    As a final remark we note that our dark matter candidate means
that gravity may get modified at scales $m^{-1}\propto 0.1\mu
\rm{m}\Bigg(\frac{H_I}{10^{13}\rm{GeV}}\Bigg)^4$. This is still
about two orders of magnitude below the current experimental
bound~\cite{Will:2005va}.

\ack
We would like to thank Wessel Valkenburg and Willem Westra
for many interesting discussions and insights on the issue of NGT.
Finally we thank John Moffat for his correspondence concerning
previous work on NGT
\\
\bibliographystyle{utcaps}
\bibliography{biblio}

\providecommand{\href}[2]{#2}\begingroup\raggedright\begin{thebibliography}{10}

\bibitem{Will:2005va}
C.~M. Will, ``The confrontation between general relativity and experiment,''
\href{http://arXiv.org/abs/gr-qc/0510072}{{\tt gr-qc/0510072}}.

\bibitem{Moffat:1994hv}
J.~W. Moffat, ``Nonsymmetric gravitational theory,'' {\em Phys. Lett.} {\bf
  B355} (1995) 447--452,
\href{http://arXiv.org/abs/gr-qc/9411006}{{\tt gr-qc/9411006}}.

\bibitem{Moffat:2004bm}
J.~W. Moffat, ``Gravitational Theory, Galaxy Rotation Curves and Cosmology
  without Dark Matter,'' {\em JCAP} {\bf 0505} (2005) 003,
\href{http://arXiv.org/abs/astro-ph/0412195}{{\tt astro-ph/0412195}}.

\bibitem{Moffat:2004nw}
J.~W. Moffat, ``Modified gravitational theory as an alternative to dark energy
  and dark matter,'' (2004)
\href{http://arXiv.org/abs/astro-ph/0403266}{{\tt astro-ph/0403266}}.

\bibitem{Prokopec:2006kr}
T.~Prokopec and W.~Valkenburg, ``Antisymmetric Metric Field as Dark Matter,''
\href{http://arXiv.org/abs/astro-ph/0606315}{{\tt astro-ph/0606315}}.

\bibitem{Janssen:2006nn}
T.~Janssen and T.~Prokopec, ``Instabilities in the nonsymmetric theory of
  gravitation,'' {\em Class. Quant. Grav.} {\bf 23} (2006) 4967--4982,
\href{http://arXiv.org/abs/gr-qc/0604094}{{\tt gr-qc/0604094}}.

\bibitem{Damour:1992bt}
T.~Damour, S.~Deser, and J.~G. McCarthy, ``Nonsymmetric gravity theories:
  Inconsistencies and a cure,'' {\em Phys. Rev.} {\bf D47} (1993) 1541--1556,
\href{http://arXiv.org/abs/gr-qc/9207003}{{\tt gr-qc/9207003}}.

\bibitem{Clayton:1995yi}
M.~A. Clayton, ``Massive NGT and spherically symmetric systems,'' {\em J. Math.
  Phys.} {\bf 37} (1996) 395--420,
\href{http://arXiv.org/abs/gr-qc/9505005}{{\tt gr-qc/9505005}}.

\bibitem{Prokopec:2005fb}
T.~Prokopec and W.~Valkenburg, ``The cosmology of the nonsymmetric theory of
  gravitation,'' {\em Phys. Lett.} {\bf B636} (2006) 1--4,
\href{http://arXiv.org/abs/astro-ph/0503289}{{\tt astro-ph/0503289}}.

\bibitem{Valkenburg}
W.~Valkenburg, ``Linearized nonsymmetric metric pertubations in cosmology,''
  {\em master's thesis at the ITP of Utecht University, Available at
  http://www1.phys.uu.nl/wwwitf/teaching/thesis.htm} (2006).

\bibitem{Chamseddine:2005at}
A.~H. Chamseddine, ``Hermitian geometry and complex space-time,'' {\em Commun.
  Math. Phys.} {\bf 264} (2006) 291--302,
\href{http://arXiv.org/abs/hep-th/0503048}{{\tt hep-th/0503048}}.

\bibitem{Mann:1982}
R.~Mann and J.~Moffat, ``Ghost properties of generalized theories of
  gravitation,'' {\em Phys. Rev. D} {\bf 26} (1982) 1858.

\end{thebibliography}\endgroup

\end{document}